\documentclass[amsmath,amssymb,aps,pra,superscriptaddress,floatfix,twocolumn,showpacs]{revtex4-1}

\usepackage{graphicx}
\usepackage{dcolumn}
\usepackage{bm}
\usepackage{leftidx}
\usepackage[utf8]{inputenc}
\usepackage{isotope}
\usepackage{mathtools}
\usepackage{braket}
\usepackage[alsoload=accepted]{siunitx}
\usepackage{hyperref}

\hypersetup{pdftitle={Spatial Pauli-blocking of spontaneous emission in optical lattices}}

\DeclareMathOperator{\vac}{vac}

\providecommand{\abs}[1]{\lvert#1\rvert}

\newcommand\varpm{\mathbin{\vcenter{\hbox{%
\oalign{\hfil${+}$\hfil\cr
          \noalign{\kern-.3ex}
          $({-})$\cr}%
}}}}

\begin{document}

\title{Spatial Pauli-blocking of spontaneous emission in optical lattices}

\author{R.~M.~Sandner}
\affiliation{Institute for Theoretical Physics, University of Innsbruck, A-6020 Innsbruck,
Austria}
\affiliation{Institute for Quantum Optics and Quantum Information of the Austrian Academy of Sciences, A-6020 Innsbruck, Austria}
\author{M.~M\"uller}%
\affiliation{Institute for Theoretical Physics, University of Innsbruck, A-6020 Innsbruck,
Austria}
\affiliation{Institute for Quantum Optics and Quantum Information of the Austrian Academy of Sciences, A-6020 Innsbruck, Austria}
\affiliation{Departamento de F\'isica Te\'orica I, Universidad Complutense, 28040 Madrid, Spain}
\author{A.~J.~Daley}
\affiliation{Institute for Theoretical Physics, University of Innsbruck, A-6020 Innsbruck,
Austria}
\affiliation{Institute for Quantum Optics and Quantum Information of the Austrian Academy of Sciences, A-6020 Innsbruck, Austria}
\affiliation{Department of Physics and Astronomy, University of Pittsburgh, Pittsburgh, Pennsylvania 15260, USA}
\author{P.~Zoller}
\affiliation{Institute for Theoretical Physics, University of Innsbruck, A-6020 Innsbruck,
Austria}
\affiliation{Institute for Quantum Optics and Quantum Information of the Austrian Academy of Sciences, A-6020 Innsbruck, Austria}

\date{July 18, 2011}

\begin{abstract}
Spontaneous emission by an excited fermionic atom can be suppressed  due to the Pauli
exclusion principle if the relevant final states after the decay are already occupied by
identical atoms in the ground state. Here we discuss a setup where a single atom is
prepared in the first excited state on a single site of an optical lattice under
conditions of very tight trapping. We investigate these phenomena in the context of two
experimental realizations: (1) with alkali atoms, where the decay rate of the excited
state is large and (2) with alkaline earth-like atoms, where the decay rate from
metastable states can be tuned in experiments. This phenomenon has potential applications
towards reservoir engineering and dissipative many-body state preparation in an optical
lattice.
\end{abstract}

\pacs{03.75.Ss, 37.10.Jk, 42.50.Ct, 67.85.-d}
\maketitle



\section{\label{sec:introduction}Introduction}
An atom in free space prepared in an excited electronic state will decay 
by spontaneous emission to lower lying electronic states. Spontaneous 
emission is a basic ingredient in light scattering from atoms, and 
processes involving spontaneous emission constitute fundamental decay, 
and thus decoherence mechanisms, when manipulating atoms with laser 
light. There is an extensive literature on suppression of the (free 
space) spontaneous emission rate. This can be achieved first of all by 
engineering the density of states of the radiation modes of the emitted 
photons, e.g. by placing emitters in a cavity \cite{Kleppner1981}, in a photonic bandgap 
material \cite{Noda2007}, or close to a surface \cite{Jhe1987}. On the other hand, recent 
experimental advances with quantum degenerate Fermi gases have motivated 
theoretical studies of ``Pauli-blocking'' spontaneous emission, where the 
spontaneous emission of an atom from the first excited electronic state 
to the ground state is blocked by other fermionic atoms occupying 
possible final states into which the atom can decay. This scenario and 
prospects for experimental observation have so far been discussed for a 
single excited atom above a Fermi sea of {\em many} trapped ground state 
atoms \cite{Busch1998,Shuve2010}, where the condition for Pauli-blocking spontaneous emission is 
$E_F > E_R$ with $E_F$ the Fermi energy of ground state atoms, and 
$E_R=\hbar^2k^2/2M$ the recoil energy, where $k$ denotes the wavenumber 
of the emitted photons and $M$ the atomic mass. 

Here we will instead discuss a complementary and conceptually simpler setup, in which 
a single atom 
is prepared in the first excited state on a single site of an optical 
lattice under conditions of very tight trapping realizing the Lamb Dicke 
limit, i.e.~$\eta = 2 \pi x_0 /\lambda_L\ll1$ with $x_0$ the size of the 
atomic wavepacket and $\lambda_L$ the wavelength of the trapping laser. 
In this case under suitable conditions, which we discuss below, even a 
{\em single} fermionic atom prepared on the same lattice site in the 
electronic ground state and motional ground state matching the atom in 
the excited state can block the dominant decay channel for the excited 
state to the ground state. Below we will analyze this suppression of 
decay, in particular in light of the new opportunities opened by recent 
experiments and theoretical proposals involving fermionic alkaline earth atoms 
\cite{Ye2008,Derevianko2011,Daley2008,Fukuhara2009,Cazalilla2009,Gorshkov2010,Swallows2011,Daley2011}, 
but also with alkali 
atoms in optical lattices \cite{Bloch2008,Lewenstein2007,Jaksch2005}. 

The paper is organized as follows. In Sec.~\ref{sec:overview} we give a brief qualitative overview of the
atomic decay dynamics and introduce two experimental schemes to observe Pauli-blocked
spontaneous emission. Details of these scenarios are discussed in Sec.~\ref{sec:details}.
In Sec.~\ref{sec:wavepacket} we
study the properties of the emitted photon wavepacket and the possibility to shape its
form by preparing the blocking atom in appropriate initial states. We conclude with a
summary in Sec.~\ref{sec:summary} and also give an outlook towards possible applications of Pauli-blocked
spontaneous emission as a tool for reservoir engineering in the context of dissipative
preparation of many-body quantum states and phases of fermions in optical lattices
\cite{Kraus2008,Diehl2008,Verstraete2009,Weimer2010,Cho2011,Diehl2011,Krauter2010,Barreiro2011,Diehl2010}.

\section{Overview}\label{sec:overview}
\begin{figure}[tbp]
	\includegraphics{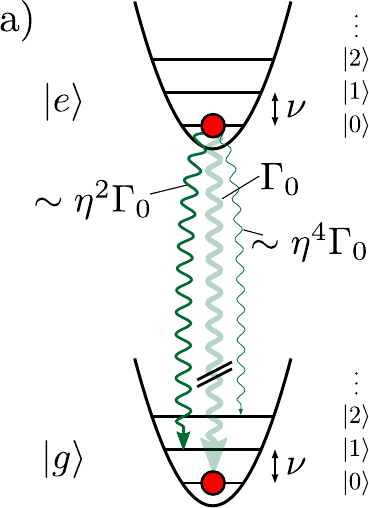}\hspace{0.8cm}
	\includegraphics{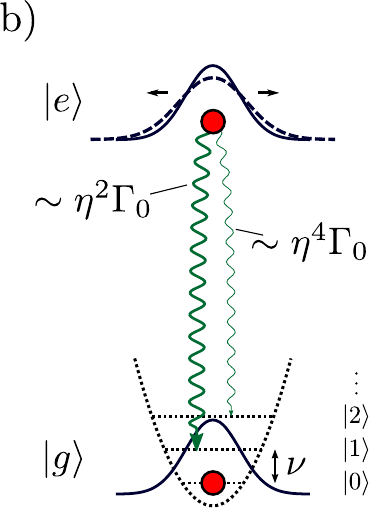}
	\caption{(a) $\Gamma_{\text{eff}}\ll \nu$: 
	The two atoms prepared in the internal excited state $\ket{e}$ and the internal ground state
	$\ket{g}$ feel the same external harmonic oscillator potential. In the Lamb-Dicke
	limit ($\eta\ll 1$), the dominant decay channel for a single particle 
	$\ket{e}\ket{0}\rightarrow \ket{g}\ket{0}$ with a rate $\Gamma_0$ is now blocked by the
	ground state atom due to the Pauli exclusion principle. The excited atom decays under
	change of its motional state with a rate $\Gamma_{\text{eff}}$ that is reduced of order $\eta^2$.  
	(b) $\Gamma_{\text{eff}}\gg \nu$: In this scenario, the notion of motional eigenstates
	is meaningless for atoms in $\ket{e}$. Yet again, both atoms can be prepared in the same
	initial motional state represented by the wavepacket $\ket{0}$ (solid line). Dispersion of the motional
	wavepacket for the excited atom (dashed line) is slow compared to the effective decay dynamics. As
	before, the dominant decay channel is blocked and the effective decay rate is reduced of
	order $\eta^2$.}
	\label{fig:fig1}
\end{figure} 
In this section, we first give a qualitative description of the effect of Pauli-blocked
spontaneous emission in optical lattices and
subsequently outline possible experimental realizations using alkali and alkaline-earth
atoms.

Let us consider a three dimensional optical lattice that is loaded with two identical fermionic atoms per
site and which is sufficiently deep that tunneling between neighboring sites is
negligible on experimental timescales. Let us further assume that on each site one of the
atoms is in an electronically excited state $\ket{e}$ while the other resides in its
internal ground state $\ket{g}$. Due to its coupling to the electromagnetic field, the
excited atom can undergo radiative decay. As a result of the Pauli exclusion principle,
the presence of the ground state atom reduces the number of available decay channels,
prolonging the lifetime of the excited state. For this blocking effect to be significant we
require that both atoms are initially prepared in the same motional state, that the atoms
are tightly confined to a region smaller than the optical wavelength of the electronic
transition (Lamb-Dicke regime) and that decay from the excited state $\ket{e}$ is only
possible to the internal ground state $\ket{g}$ (see Fig.~\ref{fig:fig1}(a)).

In the regime without tunneling the spatial potential on each lattice site is well approximated by a
three dimensional harmonic oscillator potential. Here, for simplicity
of this qualitative discussion, we regard motional excitations only in one dimension where
the oscillator trapping frequency is $\nu$ and the vibrational eigenstates are denoted
$\ket{n}$ ($n=0,1,\ldots$). This will later be extended to a 3D model. Furthermore, we
assume for the moment that atoms in the internal excited state feel the same oscillator
potential as ground state atoms, and that the natural linewidth $\Gamma$ of the state
$\ket{e}$ is small compared to $\nu$. While these requirements simplify the discussion
they are not mandatory, as we will argue below.  
For our initial state of interest, which is denoted 
$\ket{\psi_0}=c_{g0}^\dagger c_{e0}^\dagger \ket{\vac}$ in
second quantized notation, both atoms reside in the motional ground state $\ket{0}$ and
one is internally excited (see Fig.~\ref{fig:fig1}(a)). Here, $c^{\dagger}_{gn}$
($c^{\dagger}_{en}$) creates a
particle on the site we consider in the electronic state $\ket{g}$ ($\ket{e}$) and motional state
$\ket{n}$.
Treating the system within a Weisskopf-Wigner approximation
\cite{Louisell1973},
the initial state $\ket{\psi_0}$ couples
to the possible final atomic states $c_{g0}^\dagger c_{gn}^\dagger \ket{\vac}$ under
emission of a one-photon wavepacket. The resulting dynamics within this ansatz
yields an exponential decay for the population in the excited state with an effective decay rate
\begin{equation}
	\Gamma_\text{eff}=\Gamma\sum_{n\neq 0}\int_{-1}^{1}\mathrm{d}u\,N(u)
	\abs{\braket{n|\mathrm{e}^{-\mathrm{i}u\eta(a+a^\dagger)}|0}}^2.
	\label{eq:G_eff}
\end{equation}
In Eq. \eqref{eq:G_eff}, $\Gamma=d_{eg}^2\omega_0^3/(3\pi\varepsilon_0\hbar c^3)$ is the
usual single particle decay rate from $\ket{e}$ to $\ket{g}$ where $d_{eg}$ and $\omega_0$
are the dipole matrix element and frequency of the optical transition, respectively;
$a^\dagger$ ($a$) denotes the creation (annihilation) operator for the harmonic oscillator
and $N(u)$ is the angular distribution of dipole radiation projected onto the oscillator
axis. The parameter $\eta=2\pi x_0/\lambda_0$ is given by the ratio between the oscillator
ground state length $x_0=\sqrt{\hbar/(2M\nu)}$ with atomic mass $M$ and the wavelength of
the emitted light $\lambda_0=2\pi c/\omega_0$. Note that in the sum over the final
vibrational modes for the decaying atom, the mode $n=0$ is excluded due to the fermionic
property $c_{g0}^\dagger c_{g0}^\dagger=0$, which reflects the Pauli-blocking of this
particular channel. 

In the Lamb-Dicke regime $x_0\ll
\lambda_0$ the exponential in Eq.~\eqref{eq:G_eff} can be expanded in the small
parameter $\eta\ll 1$. In this limit, the leading contribution results from decay into
the vibrational mode $n=1$, yielding an effective decay rate $\Gamma_{\text{eff}}\sim\eta^2
\Gamma$. In contrast, a single particle decays at a total rate $\Gamma$. For two atoms,
the excited state's decay rate is decreased by a 
factor of order $\eta^{2}$. This results from blocking of the dominant decay channel, which
corresponds to a decay while preserving the motional state, due to 
Pauli's exclusion principle.

Let us now mention details and additional effects that modify the simplified picture
outlined so far. Coming back
to the preparation of the initial state $c_{g0}^\dagger c_{e0}^\dagger \ket{\vac}$,
we distinguish the two scenarios $\Gamma_{\text{eff}}\ll\nu$ and $\Gamma_{\text{eff}}\gg
\nu$.  
For $\Gamma_{\text{eff}}\ll \nu$ (see Fig.~\ref{fig:fig1}(a)) the decay is slow compared to the atomic motion.
Such a situation can be realized by using metastable excited states of alkaline-earth
atoms, where it is also possible to produce equal trapping potentials for ground state and
excited atoms via magic wavelength lattices \cite{Katori2003,Ye2008}.
In this sideband resolved regime it is possible to laser-excite one of the atoms while
exclusively coupling to the motional ground state.  The opposite regime
$\Gamma_{\text{eff}}\gg \nu$ (see Fig.~\ref{fig:fig1}(b)) is
naturally realized for short-lived low-lying excited states in alkali atoms. Here, laser
excitation from $\ket{g}$ to $\ket{e}$ is performed in the strong-excitation regime
\cite{Poyatos1996} where
the initial motional wavepacket remains essentially unchanged during the electronic
transition and corresponds to the motional ground state for atoms in the internal
state $\ket{g}$. In this regime a possibly different (anti-) trapping potential for excited
state atoms gives rise only to a small correction to the Pauli-blocked decay dynamics. In
this way, it is possible to initially prepare the ground and excited atom in the same
motional state for either scenario, which will therefore lead to an effective decay rate
$\Gamma_{\text{eff}}\sim\eta^2\Gamma$ in either case.

Furthermore, additional two-body effects appear, which will be 
taken into account in a more accurate treatment  based on a second quantized many particle
master equation (Section \ref{sec:details}). Since both atoms are confined to a region
much smaller than $\lambda_0$, their mutual coupling to the radiation field gives rise to
dipole-dipole interaction between the two atoms. In addition, cross-damping processes,
which in other contexts are also  responsible for super- and subradiance, can modify the
decay characteristics. These effects are more often encountered in the limit $\Gamma\gg
\nu$.
Additionally, the influence of state changing collisions between the atoms will be
addressed.

In the following two subsections we give an overview of possible realizations of
Pauli-blocked spontaneous emission with ultracold atoms in optical lattices using (A)
alkaline earth atoms and (B) alkali atoms. We will discuss the experimental requirements
and steps for each scenario and show how the electronic dynamics of such multi-level atoms
can be related to the effective two-level scheme discussed above.

\begin{figure}[tbp]
  \includegraphics[width=8cm]{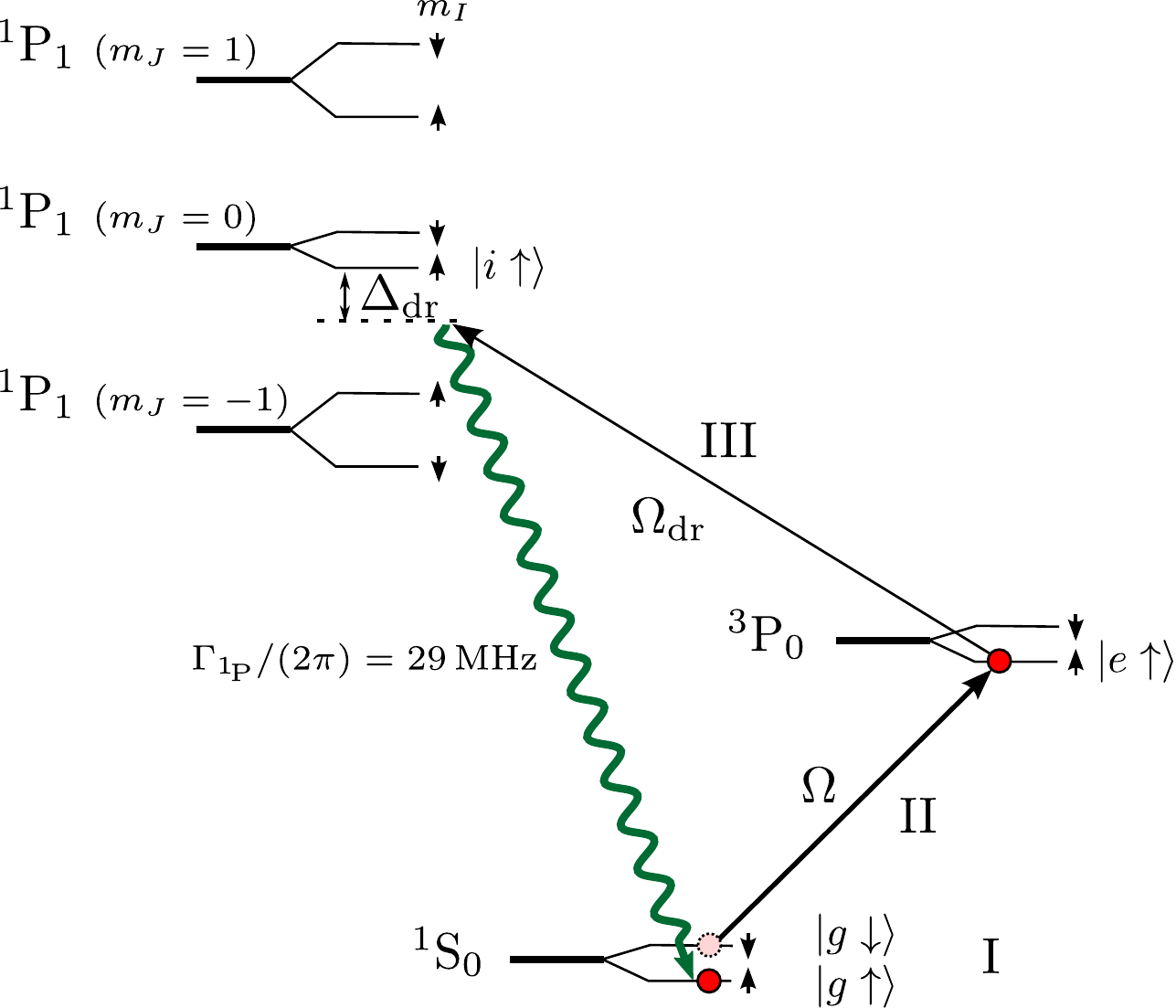}
  \caption{ Relevant level structure (not to scale) and 
	experimental sequence for alkaline
	earth-like atoms (\isotope[171]{Yb} in this example).	(I) Two atoms per lattice site are
	prepared in
	$\ket{g\uparrow}$ and $\ket{g\downarrow}$, respectively; (II) one atom is excited from
  $\ket{g\downarrow}$ to the metastable state
	$\ket{e\uparrow}$ under a flip of its nuclear spin; (III) spontaneous emission from
  $\ket{e\uparrow}$ is induced by
	admixing the fast decaying state $\ket{i\uparrow}$. Nuclear spin flips during the induced decay
	are suppressed by applying a large external magnetic field which decouples nuclear and electronic
  spin in the $^1$P$_1$ manifold. Consequently the induced decay from $\ket{e\uparrow}$ to
  $\ket{g\uparrow}$ is Pauli-blocked.
	}
  \label{fig:fig2}
\end{figure}
\begin{figure}[tbp]
	\includegraphics[width=8cm]{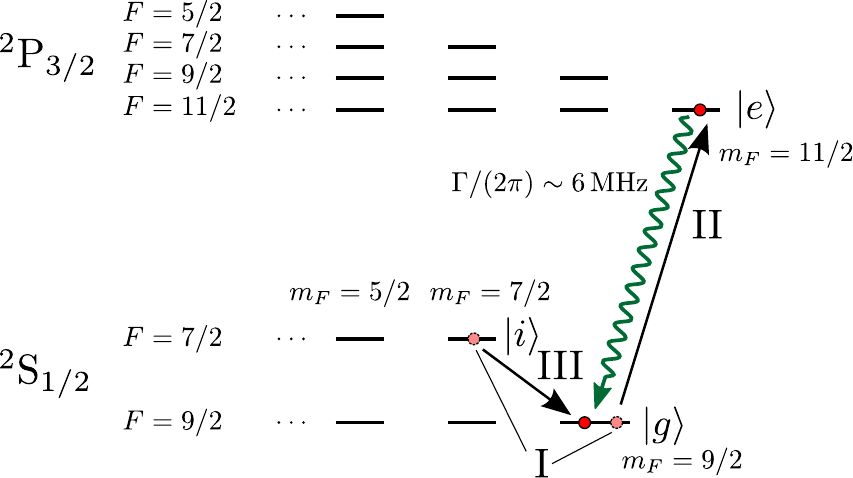}
	\caption{	Relevant level structure (not to scale) and experimental sequence for alkali
  atoms (\isotope[40]{K} in this example). 
  (I) Two atoms per lattice site are prepared in $\ket{g}$ and $\ket{i}$, respectively;
	(II) first, one
	atom is excited from $\ket{g}$ to the excited state $\ket{e}$; (III) then the other atom is transferred
	from $\ket{i}$ to $\ket{g}$. This pulse sequence is fast on the timescale given by
  $\Gamma^{-1}$. Choosing states with maximal $m_F$ values ensures that $\ket{g}$ and $\ket{e}$
  form a closed two-level system.
  The decay of the excited atom back to $\ket{g}$ is Pauli-blocked. }
	\label{fig:fig3}
\end{figure}
\subsection{Alkaline earth atoms}
Alkaline earth atoms with two valence electrons 
offer the opportunity to demonstrate the effect of Pauli-blocked
spontaneous emission in a
way that is closely related to the qualitative model outlined above. These atoms have
singlet ground states and  metastable triplet states with lifetimes on the order
of several tens of seconds. Since this exceeds accessible experimental timescales, we will
use such a state as the excited state in our scheme and induce spontaneous emission by weak laser
coupling to an intermediate state which rapidly decays to the internal ground state.
Hence, the effective
tunable decay rate can be made significantly larger than the natural linewidth of the
metastable state. By operating the 
optical lattice at the ``magic'' wavelength it is possible to have equal trapping
potentials for the ground and metastable excited state \cite{Katori2003,Ye2008}. Furthermore, the fermionic isotopes have an
additional nuclear spin degree of freedom, which we will make use of in the context of
initial state preparation.

The relevant electronic level structure and experimental sequence I through III is illustrated in 
Fig.~\ref{fig:fig2}.
(I) After
the lattice is adiabatically ramped up on an ultracold cloud of atoms, two atoms per lattice site with
the nuclear spin projections $\ket{\uparrow}$ and $\ket{\downarrow}$ reside
in their motional and internal ground state $\ket{g}\ket{0}$ forming a band insulator. (II) With a  $\pi$ pulse on 
the weakly dipole-allowed transition $\ket{g}\rightarrow \ket{e}$, one of the atoms is
transferred from $\ket{g\downarrow}\ket{0}$ to the excited state 
$\ket{e\uparrow}\ket{0}$ while flipping its nuclear spin. (III) 
Finally, spontaneous emission from $\ket{e}$ is induced by admixing a fast decaying
intermediate state. In order to avoid that the excited atom decays via the channel $\ket{e\uparrow}\rightarrow
\ket{g\downarrow}$ which is not Pauli-blocked, one can either use a sufficiently large
external magnetic field \cite{Reichenbach2007}, or alternatively choose an appropriate polarization of the
dressing laser. For both cases this leads to a decoupling of electronic and nuclear spin for the induced decay
process (see Fig.~\ref{fig:fig2}), and as the intermediate state is only 
virtually populated, $\ket{e\uparrow}$ and
$\ket{g\uparrow}$ form a closed system regarding the internal degrees of freedom.

As a consequence the described system constitutes a realization of the
model system illustrated in Fig.~\ref{fig:fig1}(a) where the excited atom can only
undergo decay under change of its motional state. The induced decay rate 
is by a factor of order $\eta^2$ smaller than it is for a single atom.
\subsection{Alkali atoms}
In contrast to the alkaline earth case, alkali atoms have one valence electron whose lower
excited states decay to the ground state on dipole-allowed transitions with a large
linewidth $\Gamma$ on the order of several $\si{\mega\hertz}$.

In Fig.~\ref{fig:fig3} 
a typical alkali level structure and a possible experimental
scheme is shown.  We choose to make use of the transition between the hyperfine sublevels
$\ket{e}$ and $\ket{g}$ with maximal $m_F$ values, because the only dipole-allowed
transition for atoms prepared in $\ket{e}$ is decay to $\ket{g}$, so these states form an effective
closed two-level system.
(I) The lattice is prepared with two atoms per lattice site in the ground state
sublevels $\ket{i}$ and $\ket{g}$ (both in their motional ground state $\ket{0}$). 
(II) The first atom is excited with a state selective $\pi$ pulse from $\ket{g}$ to
$\ket{e}$. (III) The second atom is transferred from $\ket{i}$ to 
$\ket{g}$ with a radio frequency or Raman $\pi$ pulse. The
total pulse sequence should be fast compared to the timescale $\Gamma^{-1}$ so that spontaneous emission
during the preparation sequence can be neglected. The excited atom then undergoes
radiative decay with a Pauli-blocked rate on the order of $\eta^2\Gamma$. 

As a consequence of the large linewidth $\Gamma$ imperfections are introduced that
will qualitatively modify the Pauli-blocking effect:
(i) the induced dipole-dipole interaction between atoms is strong for large $\Gamma$ and will excite
motional states that are not blocked;
(ii) due to the fast excitation pulses the transfer to $\ket{e}$ is performed in the 
strong-excitation regime where the Rabi frequency is much larger than the oscillator
spacing $\nu$ of the ground state. 
Thus, the initial motional state $\ket{0}$ is displaced by the momentum recoil
of the absorbed photon. This leads to slightly different motional states for the ground and excited
state atom; (iii) in general, the atoms encounter different (anti-)trapping potentials
for their internal ground and excited states. However, 
due to the large separation of timescales $\nu\ll \Gamma$, corrections to the dynamics due
to atomic motion during the decay are small. All these effects will be discussed in more
detail in the following section.

\section{Analysis and Results}\label{sec:details}

In this section we first discuss the many-body master equation for fermionic two-level
atoms in second quantization.  We show how Pauli-blocking emerges naturally in this
description and discuss collective effects which arise for more than one atom.
Subsequently, we present a detailed quantitative discussion of the experimental
realizations with both alkaline earth-like and alkali atoms.

\subsection{Master equation}
\label{sec:me_details}
We consider fermionic two-level atoms with internal states $\ket{e}$ and $\ket{g}$
separated by the energy $\hbar\omega_0=\hbar ck_0$ and coupled to the radiation field
acting as a bath. The center of mass motion of the atoms takes place in a three
dimensional harmonic oscillator potential with trapping frequencies
$\boldsymbol{\nu}=(\nu_1,\nu_2,\nu_3)$ for the three spatial dimensions and eigenstates
$\ket{\boldsymbol{n}}=\ket{n_1}\ket{n_2}\ket{n_3}$ ($n_j=0,1,\cdots$) on each lattice site
for a deep optical lattice with strongly suppressed tunneling.  
We are interested in a situation where the atoms
are confined to a region small compared to the optical wavelength $\lambda_0=2\pi/k_0$.
Under assumption of the Born-Markov approximation, the dynamics is described by a standard
quantum optical master equation \cite{Lehmberg1970a,Kazantsev1990,Pichler2010}
\begin{equation}
	\dot{\rho}=-\frac{\mathrm{i}}{\hbar}(H_{\text{eff}}\rho-\rho H_{\text{eff}}^\dagger)+
	\hat{\Gamma}[\rho],
	\label{eq:master}
\end{equation}
where for our purposes we explicitly take into account the indistinguishability and the
motional degrees of freedom of the atoms.  The terms appearing in the effective
non-hermitian Hamiltonian $H_{\text{eff}}$ and in the recycling term $\hat{\Gamma}[\rho]$
are discussed in this subsection. 

The effective Hamiltonian can be written as the sum of a hermitian and a non-hermitian
part, $H_{\text{eff}}=H_{\text{eff}}^{(0)}+H_{\text{eff}}^{(1)}$. In 
\begin{eqnarray}
	H_{\text{eff}}^{(0)}&=&
      \smashoperator{\sum_{\boldsymbol{n}}} \hbar
      \boldsymbol{\nu}\cdot\boldsymbol{n}\,
      \left(c_{g\boldsymbol{n}}^\dagger
      c_{g\boldsymbol{n}}+c_{e\boldsymbol{n}}^\dagger c_{e\boldsymbol{n}}\right)\nonumber\\
			&&-\smashoperator{\sum_{\boldsymbol{m}\boldsymbol{n}
      \boldsymbol{m}'\boldsymbol{n}'}}\hbar L_{\boldsymbol{n}'\boldsymbol{m}'\boldsymbol{m}\boldsymbol{n}}
			c_{e\boldsymbol{n}'}^\dagger c_{g\boldsymbol{m}'}^\dagger
			c_{g\boldsymbol{m}}c_{e\boldsymbol{n}}
	\label{eq:Heff(0)}
\end{eqnarray}
the first term accounts for the motion of the atoms in the harmonic oscillator potential.
Here $c_{\beta\boldsymbol{n}}$ ($c_{\beta\boldsymbol{n}}^\dagger$) are the annihilation
(creation) operators for an atom in the internal state $\beta\in\left\{ e,g \right\}$ and
in the vibrational mode $\ket{\boldsymbol{n}}$ of the harmonic oscillator. These operators
obey the fermionic anti-commutation relations
$\{ c_{\beta
\boldsymbol{n}}^\dagger,c_{\beta'\boldsymbol{n}'}^\dagger\}=\{ c_{\beta\boldsymbol{n}},
c_{\beta'\boldsymbol{n}'}
\}=0$ and $\{ c_{\beta
\boldsymbol{n}}^\dagger,c_{\beta'\boldsymbol{n}'}
\}=\delta_{\beta,\beta'}\delta_{\boldsymbol{n},\boldsymbol{n}'}$.
The second term in Eq.~\eqref{eq:Heff(0)}, a two particle operator, 
is the dipole-dipole interaction between the atoms, which is induced by the collective
coupling to the radiation field and has to be taken into account when the distance
between the atoms is comparable to or smaller than $\lambda_0$. It is given by
\begin{eqnarray}
	L_{\boldsymbol{n}',\boldsymbol{m}',\boldsymbol{m},\boldsymbol{n}}&=&\Gamma\smashoperator{\iint}
	\mathrm{d}^3x\,\mathrm{d}^3x'\, 
	G(k_{0}(\boldsymbol{x}-\boldsymbol{x}'))\notag\\
	&&\times\varphi_{\boldsymbol{n}'}(\boldsymbol{x})\varphi_{\boldsymbol{m}'}(\boldsymbol{x}')
  \varphi_{\boldsymbol{m}}(\boldsymbol{x})\varphi_{\boldsymbol{n}}(\boldsymbol{x}')
	\label{eq:dipole-dipole}
	\\
  G(\boldsymbol{\xi})&=&\frac{3}{4}\frac{1-3(\hat{\boldsymbol{d}}_{eg}\cdot
	\hat{\boldsymbol{\xi}})^2}{\xi^3}\label{eq:G},
\end{eqnarray}
where $\Gamma$ is the single particle decay rate from $\ket{e}$ to $\ket{g}$ as
defined above, $\varphi_{\boldsymbol{m}}(\boldsymbol{x})$ are the position space harmonic
oscillator eigenfunctions and $\hat{\boldsymbol{d}}_{eg}$ denotes the dipole matrix
element unit vector of the transition. The representation Eq.~\eqref{eq:G} for the 
function $G(\boldsymbol{\xi})$ is valid in the
limit $\abs{\xi}\ll 1$ which is fulfilled in the Lamb-Dicke regime. For broader traps with
interatomic distances $k_0\abs{\boldsymbol{x}-\boldsymbol{x}'}\sim 1$ there are corrections to
$G(\boldsymbol{\xi})$ of order $1/\xi^2$ \cite{Lehmberg1970a}.

The non-hermitian part of the effective Hamiltonian reads
\begin{eqnarray}
	H_{\text{eff}}^{(1)}&=&-\frac{\mathrm{i}}{2}\hbar\Gamma
	\sum_{\boldsymbol{n}}c_{e\boldsymbol{n}}^\dagger c_{e\boldsymbol{n}}\notag\\
	&&+\frac{\mathrm{i}}{2}\hbar \Gamma
	\smashoperator{\sum_{\boldsymbol{m}\boldsymbol{n}\boldsymbol{m}'\boldsymbol{n}'}}
	\tilde{R}_{\boldsymbol{n}'\boldsymbol{m}\boldsymbol{m}'\boldsymbol{n}}c_{e\boldsymbol{n}'}^\dagger
	c_{g\boldsymbol{m}'}^\dagger c_{g\boldsymbol{m}}c_{e\boldsymbol{n}} \label{eq:Heff(1)}\\
	\tilde{R}_{\boldsymbol{n}'\boldsymbol{m}\boldsymbol{m}'\boldsymbol{n}}
	&=&\int\mathrm{d}\Omega_{\hat{\boldsymbol{k}}}
	\,N(\hat{\boldsymbol{k}})R^\ast_{\boldsymbol{n}'\boldsymbol{m}}(\hat{\boldsymbol{k}})
	R_{\boldsymbol{m}'\boldsymbol{n}}(\hat{\boldsymbol{k}})\label{eq:me_Rcoefficients},
\end{eqnarray}
where
\begin{equation*}
  N(\hat{\boldsymbol{k}})=\frac{3}{8\pi}\left(1-\abs{\hat{\boldsymbol{d}}_{eg}\cdot
	\hat{\boldsymbol{k}}}^2\right)
\end{equation*}
is the angular distribution of dipole radiation, 
$\hat{\boldsymbol{k}}=(\hat{k}_1,\hat{k}_2,\hat{k}_3)$ denotes the unit
vector of the photon emission direction,
$R_{\boldsymbol{m}\boldsymbol{n}}(\hat{\boldsymbol{k}})\equiv\prod_{j=1,2,3}R_{m_jn_j}(\hat{k}_j)=
\prod_{j}\braket{m_j|\mathrm{e}^{-\mathrm{i}\hat{k}_j\eta_j(a_j+a_j^\dagger)}|n_j}$
the recoil matrix elements (see Appendix \ref{sec:recoil}), $\eta_j=k_0\sqrt{\hbar/(2M\nu_j)}$ the Lamb-Dicke parameter
corresponding to the $j$-axis, and $a_j$ ($a_j^\dagger$) the
annihilation (creation) operators for vibrational excitations along the $j$-axis. The
effective Hamiltonian Eq.~\eqref{eq:Heff(0)} and \eqref{eq:Heff(1)} is valid under the
assumption that the system size is small compared to the distance which light travels on
typical timescales of the system dynamics. This is always fulfilled for optical
transitions and a system size on the order of an optical wavelength.

The recycling term is given by
\begin{eqnarray}
	\hat{\Gamma}[\rho]&=&\Gamma
	\smashoperator{\sum_{\boldsymbol{m}\boldsymbol{n}\boldsymbol{m}'\boldsymbol{n}'}}
	\tilde{R}_{\boldsymbol{n}'\boldsymbol{m}'\boldsymbol{m}\boldsymbol{n}}c_{g\boldsymbol{m}}^\dagger
	c_{e\boldsymbol{n}}\rho c_{e\boldsymbol{n}'}^\dagger c_{g\boldsymbol{m}'}
	\label{eq:recycling}
\end{eqnarray}
The coefficients
$\tilde{R}_{\boldsymbol{n}'\boldsymbol{m}'\boldsymbol{m}\boldsymbol{n}}$
account for the momentum recoil of
the emitted photon during a quantum jump.

For our initial state of interest with two atoms, both residing in the ground state of the
three dimensional harmonic oscillator and one being internally excited, the density operator is given by
$\rho_0=c_{g\boldsymbol{0}}^\dagger c_{e\boldsymbol{0}}^\dagger
\ket{\vac}\bra{\vac}c_{e \boldsymbol{0}}c_{g \boldsymbol{0}}$. 
In the recycling term for this initial state $\hat{\Gamma}[\rho_0]$, quantum jump terms
that correspond to a decay of the excited atom while remaining in the ground state of the
harmonic oscillator (i.e.~terms with $\boldsymbol{m}=\boldsymbol{0}$,
$\boldsymbol{m}'=\boldsymbol{0}$) are identically zero due to the fermionic properties
$c_{g\boldsymbol{0}}^\dagger
c_{g\boldsymbol{0}}^\dagger=c_{g\boldsymbol{0}}c_{g\boldsymbol{0}}=0$, reflecting the
Pauli-blocking of this particular channel. The total initial
decay rate $\Gamma_{\text{eff}}$ of the excited atom is given by the sum over all available decay channels
\begin{align}
	\Gamma_{\text{eff}}&=\smashoperator{\sum_{\boldsymbol{m}\neq \boldsymbol{0}}}
	\braket{g\boldsymbol{0};g\boldsymbol{m}|\hat{\Gamma}[\rho_0]|g\boldsymbol{0};g\boldsymbol{m}}\notag\\
  &=\Gamma\int\mathrm{d}\Omega_{\hat{\boldsymbol{k}}}\,N(\hat{\boldsymbol{k}})
  \smashoperator{\sum_{\boldsymbol{m}\neq\boldsymbol{0}}}\abs{R_{\boldsymbol{m}\boldsymbol{0}}(\hat{\boldsymbol{k}})}^2,
  \label{eq:init_decay}
\end{align}
where $\ket{g\boldsymbol{0};g\boldsymbol{m}}=c_{g\boldsymbol{0}}^\dagger
c_{g\boldsymbol{m}}^\dagger \ket{\vac}$. Equation \eqref{eq:init_decay} is the
generalization of Eq.~\eqref{eq:G_eff} to a 3D setup. In the Lamb-Dicke limit
$\eta_j\ll1$, to second order in $\eta_j$ only decay to the three oscillator
states with one motional quantum ($\ket{g\boldsymbol{0};g\boldsymbol{m}}$ with 
$\abs{m}=1$) contributes to $\Gamma_{\text{eff}}$. Expanding the recoil
matrix elements in the small parameters $\eta_j$ yields
$\abs{R_{10}(\hat{\boldsymbol{k}})}^2=\hat{k}_j^2\eta_j^2+\mathcal{O}(\eta_j^4)$ for the
$j$-axis.  The
total initial decay rate in this limit is
$\Gamma_{\text{eff}}=\sum_{j=1,2,3}\alpha_j\eta_j^2\Gamma+\mathcal{O}(\eta_j^4)$, where
$\alpha_j$ are numerical coefficients of order $1$ and depend on the relative
orientation of the transition dipole matrix element to the $j$-axis. More details on 
these coefficients and on $\Gamma_{\text{eff}}$ beyond the Lamb-Dicke limit can be
found in appendix \ref{sec:recoil}.  For the special case of an isotropic harmonic
oscillator (i.e.~$\nu_1=\nu_2=\nu_3$) the particularly simple result
$\Gamma_{\text{eff}}=\eta^2\Gamma+\mathcal{O}(\eta^4)$ is found.

Let us now return to the discussion of the collective effects introduced by the
dipole-dipole interaction (Eq.~\eqref{eq:dipole-dipole}) and the cross-damping terms
(Eq.~\eqref{eq:Heff(1)}), still concentrating on the case of two atoms trapped in the
Lamb-Dicke regime. The dipole-dipole interaction couples different
vibrational levels of two-particle states where one atom is internally excited and the
other is in its internal ground state. In evaluating the matrix elements in Eq.
\eqref{eq:dipole-dipole}, divergences at short relative distances 
appear, which reflect the fact that, in principle, one
would have to solve the exact two-atom problem to capture the right behaviour. Effects of the 
short-range potential are important at distances comparable to a few tens of Bohr radii \cite{Pichler2010}. However, in order to estimate
the order of magnitude for the leading terms it is sufficient to treat the atoms as independent
particles being described by their internal state and center of mass wavefunction.  The
divergence of the dipole-dipole matrix elements can then be
avoided in a standard way by introducing a spatial cutoff for the relative
distance between the atoms, which yields a finite result independent of the cutoff.
By this procedure, the leading terms of the dipole-dipole matrix elements
are found to be on the order of $\Gamma/(100\eta^3)$. Note that the classical scaling of the
dipole-dipole interaction with the inverse distance cubed is reflected in the $\eta$
dependence. Another process in this case leading to dissipative redistribution between excited oscillator states
are non-diagonal cross-damping terms contained in the non-hermitian Hamiltonian Eq.
\eqref{eq:Heff(1)}. The leading terms are of order $\eta^2\Gamma$.

In the case of an experimental realization with alkali atoms where the linewidth
$\Gamma$ is large, the dipole-dipole interaction is the dominant
source of imperfections and competes with the Pauli-blocking effect as it couples 
the motional state $\rho_0$ to motional states that are not
blocked.  A compromise has to be found for $\eta$: smaller values of $\eta$ lead to a
larger blocking effect, but also to a larger dipole-dipole interaction. 
When estimating the rate at which the system leaves the initial Pauli-blocked
state due to dipole-dipole interaction one must take care of divergences in the sum over
all intermediate states. In situations where this rate is on the order of the original
$\Gamma$ even for large $\eta$, the Pauli-blocking of spontaneous emissions would no longer
be observable because the atoms will be transferred to a configuration of motional states that is 
unblocked on timescales comparable to the original spontaneous emission rate. 

In contrast, for an implementation with alkaline-earth atoms the small natural linewidth of the metastable
excited state makes the dipole-dipole interaction and the cross-damping processes negligible.
Additionally, as will be argued in the next subsection, the impact of these effects on the
spontaneous emission process which is induced by admixing an excited state with a
large decay rate can be avoided by coupling far off-resonantly to this
intermediate state.

\subsection{Alkaline earth-like atoms}\label{sec:details_ae}
In this subsection we further elaborate on the experimental realization of Pauli-blocked
spontaneous emission with alkaline earth-like atoms, commenting in more detail on the
initial state preparation in a ``magic'' wavelength optical lattice, quenching of the
metastable excited state to induce spontaneous emission and the application of an external
magnetic field to decouple the nuclear and electronic spin degrees of freedom. To be specific
we present the experimental scheme choosing the isotope \isotope[171]{Yb}, which due to
its nuclear spin $I=1/2$ exhibits a particularly simple level structure.  
The discussion, though, can be readily adapted to other fermionic alkaline
earth(-like) species, such as \isotope[87]{Sr}.

In Fig.~\ref{fig:fig2} the relevant electronic states and experimental steps
are illustrated. The doubly forbidden ``clock'' transition
$(6s^2)^1\text{S}_0\leftrightarrow(6s6p)^3\text{P}_0$ is weakly dipole-allowed due to
hyperfine mixing with higher lying P-states and has a natural linewidth
$\gamma/(2\pi)\sim \SI{10}{\milli\hertz}$ \cite{Boyd2007,Porsev2004}. The nuclear spin decouples from the electronic
degrees of freedom for the ground state with total electronic angular momentum $J=0$, and
the two magnetic sublevels are given by $\ket{^1 \text{S}_{0};m_I=\pm 1/2}$.
For the metastable excited state
$^3$P$_0$ the total angular momentum projection $m_F$ rather than the nuclear spin
projection $m_I$ is a good quantum number due to the small hyperfine
admixture of electronic angular momentum. Nevertheless, the magnetic sublevels are still almost pure $m_I$
eigenstates. The states with $m_I=\pm 1/2$ are denoted as $\ket{\uparrow}$ and
$\ket{\downarrow}$, respectively.

We choose the ground state $\ket{g\uparrow}=\ket{^1\text{S}_0;m_I=1/2}$ and the metastable
excited state $\ket{e\uparrow}=\ket{^3\text{P}_0;m_I=1/2}$ as the starting point from
which the Pauli-blocking effect can be observed (see Fig.~\ref{fig:fig1}(a)), since the
tiny natural linewidth $\gamma$ of the ``clock'' transition renders both direct radiative 
decay and dipole-dipole interaction between these two states negligible.

The requirement of equal trapping potentials for the states $\ket{e\uparrow}$ and
$\ket{g\uparrow}$ can be
met by operating the optical lattice at the ``magic'' wavelength of \SI{759}{\nano\metre}
\cite{Porsev2004,Kohno2009}. 
At this wavelength, the AC polarizability and, thus, the light shift
caused by the lattice laser is the same for the $^1$S$_0$ and the $^3$P$_0$ state to first
order in the laser intensity. In the regime of a deep optical
lattice where tunneling between the sites is negligible, the spatial potential on each
lattice site is well approximated by a 3D harmonic oscillator with trapping frequencies
$\boldsymbol{\nu}$. Vibrational frequencies of $\nu/(2\pi)=\SI{90}{\kilo\hertz}$ have
already been realized in a 1D optical lattice \citep{Barber2006}.

Let us now turn to the discussion of state preparation and the required sequence of laser
pulses.

(I) Loading of the optical lattice with two atoms per site can be achieved by
adiabatically ramping up the lattice potential on a cloud of ultracold atoms
\cite{Viverit2004,Kohl2005}. For a temperature lower than the motional level spacing, 
a band insulator with two atoms per lattice site in the motional ground state
$\ket{\boldsymbol{0}}$ and with opposite nuclear spin projections
$\ket{\uparrow}$ and $\ket{\downarrow}$ will form. The resulting state on each
lattice site is $c_{g\uparrow\boldsymbol{0}}^\dagger c_{g\downarrow
\boldsymbol{0}}^\dagger \ket{\vac}$ (see Fig.~\ref{fig:fig2} I).

(II) To prepare a suitable initial state in order to observe Pauli-blocking, the atom in
$\ket{g\downarrow}\ket{\boldsymbol{0}}$ is excited with a laser $\pi$-pulse on the
``clock'' transition with Rabi frequency $\Omega$. Recoil-free excitation is possible in
the easily accessible regime $\nu_{\text{max}}\gg\Omega\gg\gamma$ where most conveniently
the laser direction is chosen along the axis of strongest confinement. In this limit,
spontaneous emission can be neglected and the vibrational sidebands are well resolved. Choosing
$\sigma^+$ polarized light flips the nuclear spin due to hyperfine interaction and
furthermore ensures that the other atom is not affected by the pulse.   By tuning the
laser on the carrier transition,  the initial state of interest $c_{g\uparrow
\boldsymbol{0}}^\dagger c_{e\uparrow \boldsymbol{0}}^\dagger \ket{\vac}$ is
prepared. 

(III) Quenching the excited state offers the possibility to induce a tunable effective
decay rate from $\ket{e\uparrow}$ to $\ket{g\uparrow}$ that exceeds the natural linewidth
$\gamma$, and also to produce a decay that will not flip the nuclear spin state (which is
not the case for intrinsic decay of the $^3$P$_0$ manifold, as the transition is weakly
allowed due to hyperfine coupling). A dressing laser couples the  metastable excited state $\ket{e\uparrow}$ on the
weakly magnetic dipole-allowed transition to the intermediate state
$(6s6p)^1\text{P}_1$ \cite{Santra2005}, which has a large linewidth $\Gamma_{^1\!\text{P}}/(2\pi)=\SI{29}{\mega\hertz}$.
Decay from $^1$P$_1$ to states other than the $^1$S$_0$ ground state occurs with a
negligible probability.

One has to ensure that during the decay induced by this dressing the nuclear spin of the
initially excited atom is not flipped due to hyperfine interaction, 
as the decay channel $\ket{e\uparrow}\rightarrow \ket{g\downarrow}$ is not
Pauli-blocked. This can be achieved by dressing $\ket{e\uparrow}$ with a state in the
$^1$P$_1$ manifold that only contains the $\ket{\uparrow}$ component, i.e. a product state
of the form $\ket{m_J;m_I=1/2}$. The only decay channel for such a state is into
$\ket{g\uparrow}$ as follows from the selection rule $\Delta m_I=0$ for
electric dipole transitions. In the following we will discuss how such coupling to an
$\ket{\uparrow}$ state can be accomplished by either (i) using $\sigma^+$
polarized dressing light to couple $\ket{e\uparrow}$ exclusively to the magnetic sublevel
with maximal $m_F$ which is a $m_I=1/2$ eigenstate, or alternatively (ii) using $\pi$ or $\sigma^-$
polarized dressing light while decoupling nuclear and electronic spin in the
$^1$P$_1$ manifold with an external magnetic field in the Paschen-Back regime
\cite{Reichenbach2007}.
To this end we diagonalize the Hamiltonian governing the $^1$P$_1$ subspace including
hyperfine- and Zeeman interaction
\begin{equation}
  \hat{H}=A \hat{\boldsymbol{\mathrm{I}}}\cdot
  \hat{\boldsymbol{\mathrm{J}}}/\hbar^2+g_j\mu_B
  \hat{\boldsymbol{\mathrm{J}}}\cdot \boldsymbol{B}/\hbar-g_I \mu_N \hat{\boldsymbol{\mathrm{I}}}\cdot
  \boldsymbol{B}/\hbar.\label{eq:zeeman}
\end{equation}
Here $g_J$ \cite{Baumann1968} and $g_I$ \cite{Olschewski1967} are the electron and nuclear
$g$ factors, $A$ \cite{Berends1992} is the magnetic hyperfine constant and $\boldsymbol{B}$ an external
magnetic field. 
\begin{figure}[tbp]
	\begin{center}
		\includegraphics[width=\columnwidth]{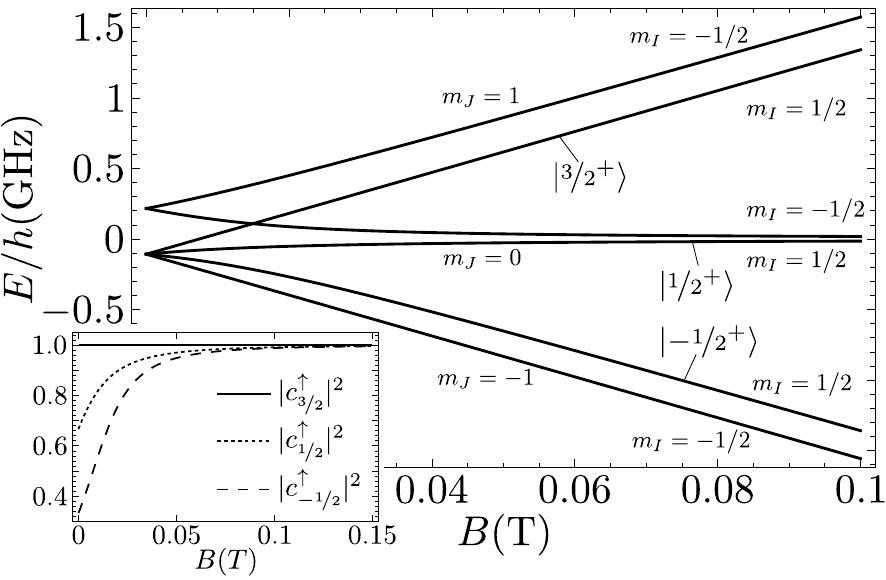}
	\end{center}
	\caption{Zeeman diagram of the $^1$P$_1$ manifold. The eigenstates approach product
	states of electronic and nuclear spin for a large magnetic field. The states suitable for
  dressing the metastable excited state $\ket{e\uparrow}$ without flipping the nuclear
  spin during the decay are $\ket{m_F^+}$ with $m_F=3/2,1/2,-1/2$. In the inset, the
  probability for decaying without nuclear spin flip is plotted as a function of the external
  magnetic field.}
	\label{fig:fig4}
\end{figure}
The eigenstates, which are denoted $\ket{m_F^\pm}$, approach product states of electronic
and nuclear spin for large values of $B$ in the Paschen-Back regime $\mu_B B\gg A$ as
illustrated in the Zeeman diagram in Fig.~\ref{fig:fig4}. The sign $\pm$ labels the
nuclear spin projection in this limit, i.e. $\ket{m_F^\pm}\rightarrow
\ket{m_J{=}m_F{\mp}1/2;m_I{=}{\pm}1/2}$. One can couple to the three states $\ket{m_F^+}$
with $m_F=3/2,1/2,-1/2$, which are suitable as intermediate states for the quenching
process, with $\sigma^+$, $\pi$ and $\sigma^-$ polarized laser light, respectively. Their
expansion in terms of product states $\ket{m_j;m_I}$ is
\begin{multline*}
  \ket{m_F^+}=c_{m_F}^{\uparrow}\ket{m_J=m_F-1/2;m_I=1/2}\\+c_{m_F}^{\downarrow}
  \ket{m_J=m_F+1/2;m_I=-1/2}.
\end{multline*}
The $B$-dependence of the coefficients is given by
\begin{align*}
	c_{m_F}^{\uparrow (\downarrow)}&=\left[ \frac{1}{2}\left( 1\varpm\frac{x+\frac{2}{3}m_F}
	{\sqrt{x^2+\frac{4}{3}m_Fx+1}} \right) \right]^{\frac{1}{2}},
\end{align*}
where $x=2(g_j\mu_B+g_I\mu_N)B/(3\abs{A})$ is a dimensionless variable for the magnetic
field strength. The
probability for decaying from $\ket{m_F^+}$ to $\ket{g\uparrow}$ or $\ket{g\downarrow}$ is
$\abs{c_{m_F}^\uparrow}^2$ and $\abs{c_{m_F}^\downarrow}^2$, respectively (see inset
Fig.~\ref{fig:fig4}). Note that (i) if one chooses to work with a $\sigma^+$
polarized dressing laser, a weak magnetic field defining the quantization axis is
sufficient because $\abs{c_{3/2}^\uparrow}^2$ is identically $1$ for arbitrary $B$-values,
and $\sigma^+$ polarized light couples $\ket{e\uparrow}$ exclusively to $\ket{3/2^+}$ in
the $^1$P$_1$ manifold. On the other hand, (ii) for $\pi$ and $\sigma^-$ polarized light
the external magnetic field has to be sufficiently strong so that $\abs{c_{\pm
1/2}^\uparrow}^2$ approaches unity and the Zeeman splitting becomes large enough so that
coupling to $\ket{m_F^-}$ is negligible. For a magnetic field of $\SI{0.05}{\tesla}$
the probability to decay without nuclear spin flip is
$>\SI{95}{\percent}$. The arguments presented here for \isotope[171]{Yb} are also
applicable to other alkaline earth-like species which may have a nuclear spin $I>1/2$.
Method (i) requires that states with maximal or minimal $m_I$ and $m_F$ values are used in the
protocol whereas method (ii) is suitable for arbitrary $m_I$ sublevels. The decoupling of
nuclear and electronic spin in the limit of large $B$ also holds for isotopes with $I>1/2$
where additional quadrupole effects have to be taken into account in the Hamiltonian
Eq.~\eqref{eq:zeeman} \cite{Reichenbach2007,Boyd2007}.

After having discussed the required experimental steps let us turn to the resulting
Pauli-blocked decay dynamics. First we show that this dynamics is described by a coupled
set of effective rate equations for the populations in $\ket{e}$ and $\ket{g}$.
Subsequently we provide a suitable set of experimental parameters.

We are interested in a regime where the detuning of the dressing laser
$\Delta_{\text{dr}}$ is large compared to all the other characteristic frequencies which
govern the system dynamics: the linewidth $\Gamma_{^1\!\text{P}}$, the
dipole-dipole interaction for two atoms in $^1$S$_0$ and $^1$P$_1$,
respectively, the dressing laser Rabi frequency $\Omega_{\text{dr}}$ and the trapping
frequency $\nu$ of both the metastable excited and the ground state. Under these conditions the
intermediate state is only virtually populated and can be adiabatically eliminated. 
Furthermore, in the resulting effective decay dynamics the
ground state atom in $\ket{g\uparrow}\ket{\boldsymbol{0}}$ acts as a spectator not taking
part in the dynamics but blocking the dominant decay channel
$\ket{e\uparrow}\ket{\boldsymbol{0}}\rightarrow \ket{g\uparrow}\ket{\boldsymbol{0}}$ for the excited
atom. The corresponding dynamics in the Lamb-Dicke limit and for the special case of an
isotropic trap is described by the rate equations
\begin{align*} 
	\dot{P}_{e\uparrow}&=-\left( \left| c_{m_F}^\uparrow\right|
	^2(\eta^2+\eta_{\text{dr}}^2)+\left| c_{m_F}^\downarrow\right|^2
	\right)\Gamma P_{e\uparrow } + \mathcal{O}(\eta^4)\\
	\dot{P}_{g\uparrow}&=\left| c_{m_F}^\uparrow\right|^2(\eta^2+\eta_{\text{dr}}^2)
	\Gamma P_{e\uparrow} +\mathcal{O}(\eta^4)\\
	\dot{P}_{g\downarrow}&=\abs{c_{m_F}^\downarrow}^2\Gamma P_{e\uparrow}.
\end{align*}
Here we have introduced the effective decay rate
$\Gamma=\Omega_{\text{dr}}^2/(4\Delta_{\text{dr}}^2)\Gamma_{^1\!\text{P}}$ and the
populations of the reduced system density matrix
\begin{eqnarray*}
	P_{e\uparrow}&=&\braket{\vac|c_{e\uparrow \boldsymbol{0}}c_{g\uparrow
	\boldsymbol{0}}\rho c_{g\uparrow \boldsymbol{0}}^\dagger c_{e\uparrow
	\boldsymbol{0}}^\dagger|\vac}\\
	P_{g\uparrow}&=&\sum_{\boldsymbol{m}\neq
	\boldsymbol{0}} \braket{\vac|c_{g\uparrow \boldsymbol{m}}c_{g\uparrow
	\boldsymbol{0}}\rho c_{g\uparrow\boldsymbol{0}}^\dagger c_{g\uparrow
	\boldsymbol{m}}^\dagger|\vac}\\
	P_{g\downarrow}&=&\sum_{\boldsymbol{m}}\braket{\vac|c_{g\downarrow
	\boldsymbol{m}}c_{g\uparrow \boldsymbol{0}}\rho c_{g\uparrow \boldsymbol{0}}^\dagger
	c_{g\downarrow\boldsymbol{m}}^\dagger|\vac}.
\end{eqnarray*}
The latter describe the probability of finding
the initially excited atom in $\ket{e\uparrow}$, $\ket{g\uparrow}$ and
$\ket{g\downarrow}$, respectively, while the other atom remains in
$\ket{g\uparrow}\ket{\boldsymbol{0}}$. The two Lamb-Dicke parameters $\eta$ and
$\eta_{\text{dr}}$ correspond to the momentum recoils of the induced spontaneous emission
and the dressing laser photon, respectively. The main result for the initial state
$c_{g\uparrow\boldsymbol{0}}^\dagger c_{e\uparrow\boldsymbol{0}}^\dagger \ket{\vac}$
described by $P_{e\uparrow}=1$ and $P_{g\uparrow}=P_{g\downarrow}=0$ is a total decay rate
from $\ket{e\uparrow}\ket{\boldsymbol{0}}$ given by
$\Gamma_{\text{eff}}=(\abs{c_{m_F}^\uparrow}^2(\eta^2+\eta_{\text{dr}}^2)+
\abs{c_{m_F}^\downarrow}^2)\Gamma$.  In the regime of interest
$\abs{c_{m_F}^\uparrow}^2\rightarrow 1$ and $\abs{c_{m_F}^\downarrow}^2\rightarrow 0$, the
decay rate $\Gamma_{\text{eff}}$ is of order $\eta^2$ smaller than the effective
induced decay rate $\Gamma$ from $\ket{e\uparrow}$ to $\ket{g\uparrow}$ for a single atom.

In this discussion so far we have not included elastic and inelastic collisions between
the two atoms. Collisional shifts do not affect the experiment as long as they are small
compared to the motional level spacing. State changing collisions can lead to loss of the
atoms from the lattice. While, in principle, the collisional loss from two atoms in $^3$P$_0$ could be
strong, this situation is never encountered in our protocol. Here, we only require the
$^1$S$_0$--$^3$P$_0$ inelastic collisions to be small compared to the quenching rate in
the experiment.

A suitable choice for the detuning $\Delta_{\text{dr}}$ is dictated by the requirement
that this has to be the largest frequency scale in the dynamics and simultaneously a 
sufficiently large effective decay rate $\Gamma$ has to be induced for a given dressing
laser Rabi frequency $\Omega_{\text{dr}}$. 
For \isotope[171]{Yb} with a typical trapping frequency $\nu/(2\pi)=\SI{90}{\kilo\hertz}$,
the Lamb-Dicke parameters $\eta=0.28$ and $\eta_{\text{dr}}=0.09$ correspond to
transitions with wavelengths $\lambda=\SI{399}{\nano\metre}$ and 
$\lambda_{\text{dr}}=\SI{1285}{\nano\metre}$, respectively. 
The dipole-dipole interaction between two atoms in $^1$S$_0$ and the virtually populated
$^1$P$_{0}$, respectively, can be estimated to be on the order of
$\Gamma_{^1\!\text{P}}$ so that $\Delta_{\text{dr}}\gg \Gamma_{^1\!\text{P}}$ is required.
For $\Delta_{\text{dr}}=10\Gamma_{^1\!\text{P}}$, with a 
Rabi frequency $\Omega_{\text{dr}}=\SI{4}{\mega\hertz}$, an effective decay rate of
$\Gamma=\SI{220}{\hertz}$ can be reached. Note that as an alternative to the direct coupling
$^3$P$_0$$\leftrightarrow$$^1$P$_0$, quenching of the metastable excited state can be done 
via a two-photon Raman process involving the dipole-allowed transition 
$^3$P$_0$$\leftrightarrow$$^3$S$_1$ and the intercombination transition 
$^3$S$_1$$\leftrightarrow$$^1$P$_1$ which can lead to a larger two-photon Rabi frequency
$\Omega_{\text{dr}}$ \cite{Reichenbach2007}. 
\subsection{Alkali atoms}\label{sec:details_alkali}
Let us now
turn to alkali atoms with their fastly decaying low lying transitions (see
Fig.~\ref{fig:fig3}). Here, both the laser excitation and the subsequent decay
dynamics are fast compared to the motion of the atoms in the trap. In contrast to the alkaline
earth case, where it is important to have equal trapping potentials for ground and excited
state atoms, the notion of motional eigenstates becomes irrelevant for the spectrally very broad
excited states in alkali atoms, therefore the optical potential for internally excited atoms does not play
a crucial role in this scenario.

First we will analyze the experimental requirements and sequences to observe Pauli-blocked
spontaneous emission using alkali atoms. Subsequently, we will discuss
imperfections associated with the large linewidth of the considered transition and estimate
their impact.

The relevant level scheme and the sequence of laser pulses is sketched in
Fig.~\ref{fig:fig3} for \isotope[40]{K}  as a
representative of the alkali family, though the scheme presented here is not specific to a
particular species and can readily be adapted to other fermionic alkali isotopes. 
We propose to use the magnetic sublevels with maximal projection of
total angular momentum $m_F$, i.e. $\ket{g}=\ket{^2 \text{S}_{1/2};F=9/2; m_F=9/2}$ and
$\ket{e}=\ket{^2 \text{P}_{3/2}; F=11/2; m_F=11/2}$, as ground and excited states as
depicted in Fig.~\ref{fig:fig1}(b). For atoms in the internal state $\ket{e}$, the decay to
$\ket{g}$ is the only dipole-allowed transition, ensuring that these states form a closed
effective two-level system regarding spontaneous emission. To prepare an initial state
suitable to observe the Pauli-blocking effect, the following experimental sequence can be
used.

(I) The lattice is adiabatically ramped up on an ultracold cloud of atoms which are
internally in a $\SI{50}{\percent}$ mixture of the state $\ket{g}$ and the magnetic
ground state sublevel $\ket{i}=\ket{^2\text{S}_{1/2};F=7/2;m_F=7/2}$, which leads 
to a band insulator with two atoms per site in the internal
states $\ket{g}$ and $\ket{i}$, respectively, provided the temperature is lower then the
motional level spacing \cite{Viverit2004,Kohl2005}. For sufficiently deep lattices, when tunneling is negligible on
the experimental timescale, the state on each lattice site is $\ket{\psi_{\text{I}}}=c_{i\boldsymbol{0}}^\dagger
c_{g\boldsymbol{0}}^\dagger \ket{\vac}$.
(II) A $\pi$-pulse is applied on the transition 
$\ket{g}\rightarrow \ket{e}$ with a Rabi frequency $\Omega_1$ to internally excite one of the atoms. 
On the one hand, we require $\Omega_1\gg \Gamma$ so that spontaneous emission
during the excitation can be neglected to lowest order.
On the other hand, $\Omega_1$ has
to be small compared to the hyperfine splitting between the states $\ket{g}$ and
$\ket{i}$, so that the other atom is not affected by the pulse. As the ground state
hyperfine splitting typically is on the order of $\si{\giga\hertz}$, both requirements
can be fulfilled simultaneously.
For alkali atoms $\Omega_1 \gg \Gamma$ also implies 
$\Omega_1 \gg \nu$, thus the transfer takes place in the strong-excitation regime
\cite{Poyatos1996}. The initial motional wavepacket is essentially unchanged during the
excitation, apart from a momentum kick $\hbar \boldsymbol{k}_L$ by the absorbed photon,
where $\boldsymbol{k}_L$ is the laser wave vector. The state prepared by
the laser pulse is therefore $\ket{\psi_{\text{II}}}=\sum_{\boldsymbol{n}}
r_{\boldsymbol{n}} c_{i\boldsymbol{0}}^\dagger c_{e\boldsymbol{n}}^\dagger
\ket{\vac}$ where the expansion coefficients $r_{\boldsymbol{n}}$ are related to the recoil
momentum associated with the absorption of a laser photon:
$r_{\boldsymbol{n}}=\braket{\boldsymbol{n}|\mathrm{e}^{\mathrm{i}\boldsymbol{k}_L\cdot 
\hat{\boldsymbol{X}}}|\boldsymbol{0}}$. For
tight traps, in the Lamb-Dicke limit, the probability to find the excited
atom in its motional state $\ket{\boldsymbol{0}}$ after the laser pulse is
$\abs{r_{\boldsymbol{0}}}^2\sim
1-\sum_j \hat{k}_{L,j}^2 \eta_j^2$ in leading order in $\eta_j$.
(III) A second $\pi$-pulse transfers the other atom from $\ket{i}$ to $\ket{g}$ with a
radio frequency pulse or by a two-photon Raman transition with effective Rabi frequency
$\Omega_2\gg \Gamma$. Spontaneous emission from the
excited state is neglibigle during the whole pulse sequence when the total time needed
to perform both pulses is short compared to $\Gamma^{-1}$. 
As the energy gap between the initial and final state for the second pulse
is not an optical frequency but the hyperfine splitting of the ground state, momentum
recoil can be neglected in this case. The state prepared after the second pulse is
$\ket{\psi_{\text{III}}}=\sum_{\boldsymbol{n}} r_{\boldsymbol{n}} c_{g\boldsymbol{0}}^\dagger
c_{e\boldsymbol{n}}^\dagger \ket{\vac}$. This is the initial state for the
Pauli-blocked decay dynamics.

The time evolution of the initial system density matrix
$\rho_0=\ket{\psi_{\text{III}}}\bra{\psi_{\text{III}}}$ is
determined by the master equation Eq.~\eqref{eq:master}, for both shallow or tight
confinement in the lattice. The initial effective decay rate can be calculated in analogy
to Eq.~\eqref{eq:init_decay}, with $\ket{\psi_{\text{III}}}$ as the initial state. For the
special case of an isotropic trap in the Lamb-Dicke limit it is found to be
$\Gamma_{\text{eff}}=2\eta^2\Gamma+\mathcal{O}(\eta^4)$, where the factor 2 reflects the
fact that there is a momentum kick involved both during the initial state preparation of
$\ket{\psi_{\text{III}}}$ and during the spontaneous emission of a photon. 

The dispersion of the excited
atom's motional wave
packet due to the different optical potential for the excited state can be
described by an additional term $\sum_{\boldsymbol{n},\boldsymbol{n}'}\hbar
T^e_{\boldsymbol{n}',\boldsymbol{n}}c_{e\boldsymbol{n}'}^\dagger c_{e\boldsymbol{n}}$
in the effective Hamiltonian Eq.~\eqref{eq:Heff(0)}. The rates
$T_{\boldsymbol{n}',\boldsymbol{n}}^e$ are typically on the order of $\nu$ and can be neglected compared to the
effective decay rate.

Let us return to collective effects in the master equation. The dipole-dipole interaction
induces transitions away from the initial state towards motional states which are not
Pauli-blocked and decay essentially with the rate $\Gamma$. Therefore, it competes with
the effect one wishes to observe. A compromise has to be found regarding the confinement
strength: tighter confinement and thus smaller values of $\eta$ are preferable on the one
hand in order to suppress recoil induced change of the initial motional wavepacket, but lead to more
pronounced dipole-dipole interaction on the other hand. For the
alkali scheme where the atoms are prepared in internal states which encounter strong
dipole-dipole interaction, the rate of
transitions of the system away from Pauli-blocked states is significant. Estimating these rates
is difficult because of diverging matrix elements which emerge if the interaction
potential at short interatomic distances is not properly treated. For cutoffs of the order
of several hundred Bohr radii, which is a typical value for Potassium, simple estimates
show that the timescale for transitions away from Pauli-blocked states is on the order of
the original $\Gamma$ even for $\eta$ of order one, with shorter cutoffs leading to larger values. 
To determine whether this is realistic in experiments
would require a much more detailed treatment of the interatomic potential at
short scales. Avoiding real population of states which encounter a large dipole-dipole
interaction is an advantage of the alkaline-earth scheme compared to the alkali scheme.

\section{Shape of the emitted photon wavepacket}
\label{sec:wavepacket}
\begin{figure}[tbp]
  \begin{center}
  \includegraphics[width=\columnwidth]{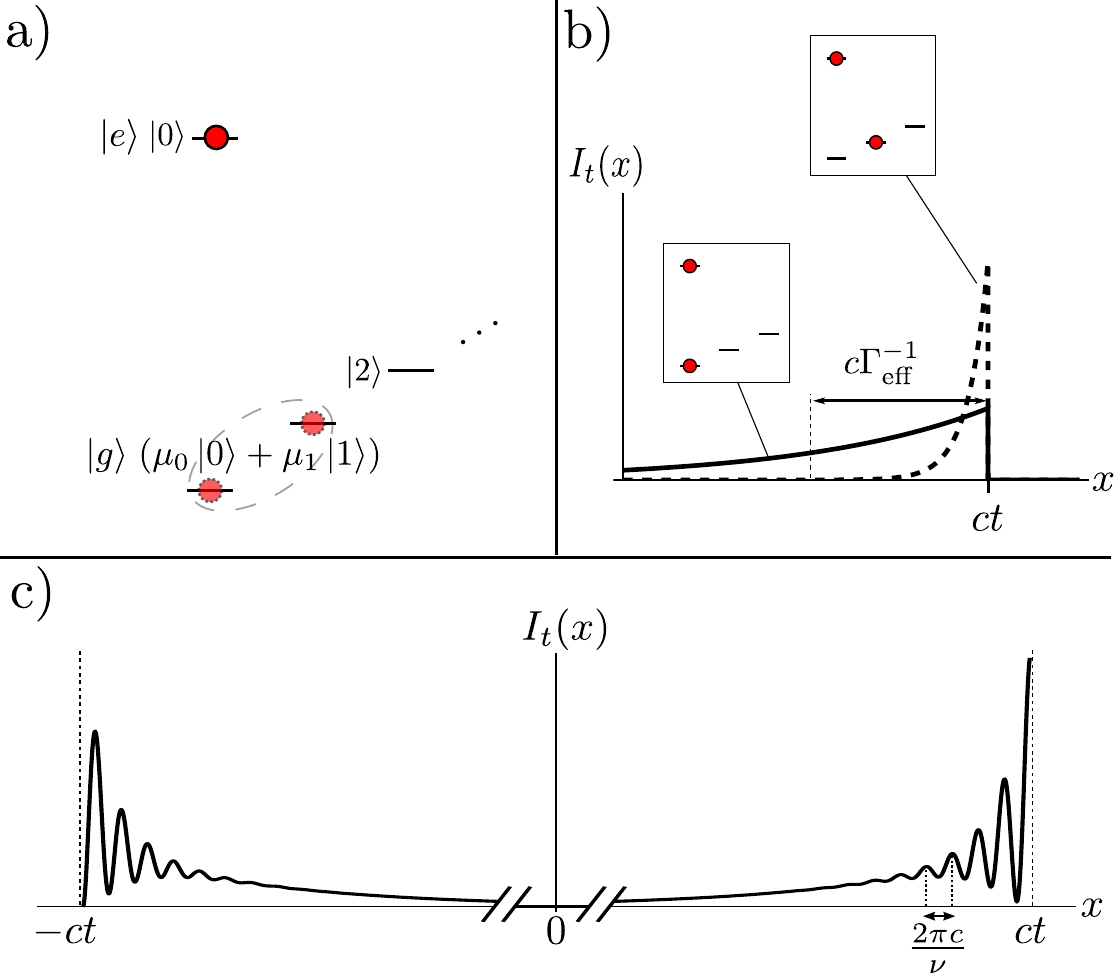}
  \end{center}
  \caption{(a) Energy diagram of the initial state: 
	two fermions with blocking atom in a superposition of the motional states
	$\ket{0}$ and $\ket{1}$. (b) Spatial intensity distribution $I_t(x)$ of the emitted
	photon wavepacket at time $t$ for the limiting cases where the blocking atom is in
	$\ket{0}$ or $\ket{1}$, respectively. In both cases, the outgoing wavepacket
	propagates with the speed of light and falls off
	exponentially with a width $c/((1-\alpha\eta^2)\Gamma)$ ($c/(\alpha\eta^2\Gamma)$) for
	the initial state $\ket{g1;e0}$ ($\ket{g0;e0}$). (c) The blocking atom is prepared in a superposition of
  motional states, we choose $\mu_0=1-\eta^2/2$ and $\mu_1=\mathrm{i}\sqrt{1-\mu_0^2}$.
  As this is the motional state which the excited atom would reach by sending out a photon along the
  negative $\hat{\boldsymbol{x}}$-axis, emission in this direction is suppressed at the
  beginning of the atomic decay ($t=0$).
  Hence, the photon intensity is zero at $x=-ct$ and has a maximum at $x=+ct$. During the
  decay of the excited atom, the blocking atom oscillates in the trap with a frequency $\nu$, which
  imprints the spatial period $2\pi c/\nu$ onto the photon wavepacket.}
  \label{fig:fig5}
\end{figure}
In the previous sections we have discussed the atomic dynamics of the Pauli-blocked
decay. In addition, this effect also becomes manifest
in the intensity distribution of the emitted photon wavepacket. 
Compared to the decay in absence of a blocking atom, due to the
prolonged lifetime of the excited state, we will see a broader spatial extension of the 
wavepacket. Furthermore, to
some extent a temporal shaping of the emitted photon is possible by preparing the blocking
atom initially in a superposition of motional states. 
For simplicity, we regard the case where the
blocking atom is in a superposition of the lowest two motional states (see
Fig.~\ref{fig:fig5}(a)). This situation
captures the essential effect as the contribution of higher motional states in the
superposition leads to modifications of order $\eta^4$ or less.
The photon shaping effect can be most explicitly  observed 
in the regime $\Gamma\ll \nu$, which can be realized in the setup with alkaline earth
atoms as discussed above. In this scheme, the appropriate initial state could be prepared after step II, 
by coherently transferring part of the population of
the blocking atom from the motional state $\ket{0}$ to $\ket{1}$.

We study this model system in a Weisskopf-Wigner ansatz, where we have one excitation either
in the atomic system or in the radiation field as an one-photon wavepacket. Compared to
the previous treatment where we have traced over the radiation field, here the
information about the photon is contained. We consider motional excitations of the atoms
only in one dimension and additionally, for the excited atom, only include the lowest
motional band $\ket{0}$ as shown in Fig.~\ref{fig:fig5}(a). Therefore, we do not include
the effects of the dipole-dipole interaction and cross-damping.
As discussed above, in the limit $\Gamma\ll \nu$, which corresponds to the
alkaline-earth case, the effect of these interactions can be suppressed by choosing a
sufficiently large detuning in the quenching process.

Within this ansatz,
the state can be written as
\begin{multline}
  \ket{\psi(t)}=
				a_{0e}(t)\ket{g0;e0}+a_{1e}(t)\ket{g1;e0}\\+
	\smashoperator{\sum_{n>m,\boldsymbol{k}}}a_{mn,\boldsymbol{k}}(t)\ket{gm;gn;1_{\boldsymbol{k}}},
		\label{eq:weisskopf-wigner}
\end{multline}
where the first two terms correspond to the situation before the decay and
the last term corresponds to both atoms in their internal groundstate and an outgoing one-photon
wavepacket.
The initial condition is $a_{0e}(0)=\mu_0$, $a_{1e}(0)=\mu_1$,
$a_{mn,\boldsymbol{k}}(0)=0$ for all $m,n,\boldsymbol{k}$ and
$\abs{\mu_0}^2+\abs{\mu_1}^2=1$.
We solve the Schrödinger equation of the Hamiltonian
\begin{eqnarray*}
	H&=&
  \sum_{n}\hbar \omega_0 c_{en}^\dagger c_{en}
  +\smashoperator{\sum_{\sigma\in\left\{ e,g
  \right\},n}}\hbar \nu n c_{\sigma n}^\dagger c_{\sigma n}
	+\sum_{\boldsymbol{k}}\hbar \omega_k b_{\boldsymbol{k}}^\dagger b_{\boldsymbol{k}}\\
	&&-\Bigl(\sum_{\boldsymbol{k}}\mathcal{E}_k \boldsymbol{d}_{eg}\cdot
  \boldsymbol{e}_{\boldsymbol{k}}
  b_{\boldsymbol{k}}\sum_{n,m}R_{nm}(\hat{\boldsymbol{k}}\cdot \hat{\boldsymbol{\chi}})
	c_{en}^\dagger c_{gm}+ \text{h.c.}\Bigr),
\end{eqnarray*}
which consists of internal and center of mass degrees of freedom of the atoms, the free
radiation field and the coupling between atoms and field in dipole and rotating wave
approximation. Here, in addition to the operators and parameters defined before,
$b_{\boldsymbol{k}}$ ($b_{\boldsymbol{k}}^{\dagger}$) is the annihilation (creation)
operator for a photon with frequency $\omega_k$ and wave vector (polarization) 
$\boldsymbol{k}$ ($\boldsymbol{e}_{\boldsymbol{k}}$), $\hat{\boldsymbol{\chi}}$ is a unit vector in direction of the
1D harmonic oscillator potential and
$\mathcal{E}_k=\mathrm{i}\sqrt{\hbar\omega_k/(2\varepsilon_0L^3)}$ with quantization
volume $L^3$. 
The solution for the coefficients $a_{0e}(t)$, $a_{1e}(t)$ and $a_{mn,\boldsymbol{k}}(t)$ 
is obtained by the standard resolvent method \cite{Louisell1973} and is given in Appendix
\ref{sec:weisskopf-wigner}.  

From the spectral distribution of the photon $a_{mn,\boldsymbol{k}}(t)$ we calculate the
first order correlation function
$I(\boldsymbol{r},t)=\braket{\psi(t)|
\boldsymbol{E}^{(-)}(\boldsymbol{r})\boldsymbol{E}^{(+)}(\boldsymbol{r})|\psi(t)}$, which
is proportional to the electric field intensity. Here,
$\boldsymbol{E}^{(+)}(\boldsymbol{r})=\sum_{\boldsymbol{k}}\mathcal{E}_{k}
\boldsymbol{e}_{\boldsymbol{k}}b_{\boldsymbol{k}}\mathrm{e}^{\mathrm{i}\boldsymbol{k}\cdot
\boldsymbol{r}}$ and its hermitian conjugate $\boldsymbol{E}^{(-)}(\boldsymbol{r})$ is the positive and 
negative part of the electric field operator, respectively.
The probability of detecting a photon between the times $t$ and $t+\mathrm{d}t$ in a volume element
$r^2\mathrm{d}r\,\mathrm{d}\Omega$ around $\boldsymbol{r}$ is proportional to
$I(\boldsymbol{r},t)r^2\mathrm{d}r\,\mathrm{d}\Omega\,\mathrm{d}t$
\cite{Glauber1963}.

Let us consider the atomic system placed at the origin of our coordinate system, with 
$\boldsymbol{d}_{eg}$ parallel to the $\hat{\boldsymbol{z}}$-axis and 
$\hat{\boldsymbol{\chi}}$ parallel to $\hat{\boldsymbol{x}}$. We analyze
$I(\boldsymbol{r},t)$ in the far zone limit $k_0r\gg 1$ and for times
$t\gg \Gamma_{\text{eff}}^{-1}$, when the initially excited atom certainly has decayed to
the ground state. In the following, we discuss the properties of $I(\boldsymbol{r},t)$
along the $\hat{\boldsymbol{x}}$-axis where we denote 
$I(\boldsymbol{r},t)r^2|_{\boldsymbol{r}=(x,0,0)}\equiv I_t(x)$. The complete form of
$I(\boldsymbol{r},t)$ can be found in Appendix \ref{sec:weisskopf-wigner}.

For the two limiting cases of the initial state,
where all of the population of the blocking atom is either in the motional state
$\ket{0}$ or $\ket{1}$ (see Fig.~\ref{fig:fig5}(b)), 
the decay of the excited atom will be maximally blocked or
essentially not blocked, respectively. 
The corresponding decay rates to order $\mathcal{O}(\eta^2)$ are
$\Gamma_{\text{eff}}^{(0)}=\alpha\eta^2\Gamma$ and
$\Gamma_{\text{eff}}^{(1)}=(1-\alpha\eta^2)\Gamma$. Consequently, the outgoing exponential photon wave
packet, which has a spatial width $c/\Gamma^{(0,1)}_{\text{eff}}$ and propagates with the speed of
light $c$, is much broader for the initial state $\ket{g0;e0}$ compared to the case of an initial
state $\ket{g1;e0}$.

Let us now consider the case where the blocking atom initially is in a superposition of motional
states. As such a state is not an eigenstate of the harmonic trapping potential, the blocking
atom oscillates in the trap while the decay of the excited atom takes place. Due to this
osciallation, the preferred direction of the emitted photon varies in time.
The characteristic frequency for the dynamics in
the trap is $\nu$, which is imprinted onto the photon wavepacket as a spatial oscillation
with a period $2\pi c/\nu$ (see Fig.~\ref{fig:fig5}(c)).

We finally point out that other techniques to induce dynamics for the blocking atom could
allow for more sophisticated shaping of the emitted photon. For example, Rabi oscillations
could be driven for the blocking atom between $\ket{g0}$ and another, decoupled internal state,
or the trapping potential could be modulated during the Pauli-blocked decay.
\section{Summary and Outlook}
\label{sec:summary}
In this work we have studied how the
Pauli exclusion principle can give rise to a suppression of spontaneous emission from
electronically excited states for cold fermionic atoms stored in optical lattices.   
Here the presence of both a ground state and an excited
state atom at the same lattice site can block the dominant decay channel for the excited
atom, thereby significantly decreasing the spontaneous emission rate. Complementary to the
atomic dynamics we have studied how the decay dynamics in the presence of a blocking atom
manifests itself in the characteristics of the emitted photon. We have suggested and
analyzed experimental realizations with alkaline earth atoms and also with alkali atoms.

On the one hand, from a conceptual point of view, observing these effects for fermionic
atoms in optical lattices would be the first experimental demonstration of Pauli-blocked
spontaneous emission. On the other hand, from a more practical perspective, the use of
this blocking effect in a controllable way can constitute an additional, valuable tool in
the context of reservoir engineering in cold atom systems. Here, the idea of engineering a
controlled coupling to an environment for the dissipative preparation of entangled states
and many-body quantum phases has been explored both theoretically
\cite{Diehl2008,Verstraete2009,Kraus2008,Weimer2010,Cho2011,Diehl2011} 
and in experiments with atomic ensembles \cite{Krauter2010} and trapped ions
\cite{Barreiro2011}. 
In particular, for cold fermionic atoms in optical lattices Diehl \textit{et al.}~\cite{Diehl2010} 
have proposed and analyzed a scenario, where quasi-local
single-fermion dissipative processes can be tailored such that they lead to ``cooling''
into a BCS-type state of $d$-wave symmetry. The corresponding set of fermionic quantum
jump operators for the dissipative dynamics are suggested to be implemented via a
stroboscopic sequence of coherent and dissipative steps in a system of two-component
fermionic alkaline-earth atoms. Here, controlled induced spontaneous decay, which is
suppressed (enabled) in the presence (absence) of a second fermionic atom, as studied in
the present work, constitutes the dissipative ingredient, which also warrants the required
Fermi statistics of the quantum jump operators. 

In addition, the possibility to control to some extent the spatial and temporal emission
characteristics of the outgoing photon under induced spontaneous decay in the presence of
a suitably prepared blocking atom, might constitute an interesting tool in the context of
the development of (directed) single-photon sources \cite{Grangier2004}.

\begin{acknowledgments}
We thank Mikhail Baranov, Ivan Deutsch, Sebastian Diehl, Jun Ye and Wei Yi for stimulating discussions.
R.M.S.~thanks Miguel--Angel Mart\'in--Delgado and the
Departamento de F\'isica Te\'orica I at Universidad Complutense Madrid for hospitality. This
work was supported by the Austrian Science Foundation through SFB F40 FOQUS 
and the EU through IP AQUTE and NAMEQUAM.
\end{acknowledgments}
\appendix
\section{Recoil matrix elements and initial Pauli-blocked decay rate}\label{sec:recoil}
In Eqs.~\eqref{eq:Heff(1)} and \eqref{eq:recycling}, the recoil matrix elements
$
	R_{\boldsymbol{m}\boldsymbol{n}}(\hat{\boldsymbol{k}})\equiv\prod_{j=1,2,3}R_{m_j
	n_j}(\hat{k}_j)
$
are given by
\begin{multline*}
	R_{m_jn_j}(\hat{k}_j)=\mathrm{e}^{-\hat{k}_j^2\eta_j^2/2}
	\sqrt{\frac{\min(m_j,n_j)!}{\max(m_j,n_j)!}}(-\mathrm{i}\hat{k}_j\eta_j)^{\abs{m_j-n_j}}\notag\\
	\times L_{\min(m_j,n_j)}^{\abs{m_j-n_j}}(\hat{k}_j^2\eta_j^2),
\end{multline*}
where
\begin{align*}
	L_b^c(x)&=\sum_{j=0}^b(-1)^j \binom{b+c}{b-j}\frac{x^j}{j!}
\end{align*}
is the generalized Laguerre polynomial. For a tight trapping in the Lamb-Dicke limit, the
recoil matrix elements can be expanded in the small parameter $\eta_j\ll 1$.
To order $\eta_j^2$, the result is
\begin{align*}
  R_{m_jn_j}(\hat{k}_j)&\approx
  \begin{cases}
    1-(m+\frac{1}{2})\hat{k}_j^2\eta_j^2\ &\text{for $m_j=n_j$}\\
    -\mathrm{i}\hat{k}_j\eta_j\sqrt{m} & \text{for $\abs{m_j-n_j}=1$}\\
    -\frac{1}{2}\hat{k}_j^2\eta_j^2\sqrt{m(m-1)} & \text{for $\abs{m_j-n_j}=2$,}
  \end{cases}
\end{align*}
where we have defined $m=\max(m_j,n_j)$.
The dominant matrix elements are those with $m_j=n_j$, whereas matrix elements with
$\abs{m_j-n_j}=1$ and $\abs{m_j-n_j}=2$ are of order $\eta_j$ and $\eta_j^2$, respectively.

The master equation coefficients
$\tilde{R}_{\boldsymbol{n}'\boldsymbol{m}'\boldsymbol{m}\boldsymbol{n}}$ defined in 
Eq.~\eqref{eq:me_Rcoefficients} are real, because of parities they are zero if any of the
three components of $\boldsymbol{n}'+\boldsymbol{m}'+\boldsymbol{m}+\boldsymbol{n}$ is odd, and
they are invariant under exchange $(\boldsymbol{n}'\boldsymbol{m}')\leftrightarrow
(\boldsymbol{m}\boldsymbol{n})$ and under exchange $n'_j\leftrightarrow m'_j$ or
$m_j\leftrightarrow n_j$ for any $j=1,2,3$. In Lamb-Dicke expansion to order $\eta_j^2$,
the only contributing coefficients are those with (i) $\boldsymbol{m}=\boldsymbol{n}$ and
$\boldsymbol{m}'=\boldsymbol{n}'$, (ii) $\abs{m_j-n_j}=\abs{m'_j-n'_j}=\delta_{i,j}$ for
an $i\in\left\{ 1,2,3 \right\}$, (iii) $\boldsymbol{m}=\boldsymbol{n}$ and
$\abs{m'_j-n'_j}=2\delta_{i,j}$ for an $i\in\left\{ 1,2,3 \right\}$ (or vice versa). In
these cases, we find 
\begin{align}
	R_{\boldsymbol{n}'\boldsymbol{m}'\boldsymbol{m}\boldsymbol{n}}&\approx
	\begin{cases}
		1-\sum_j(m'_j+m_j+1)\alpha_j\eta_j^2 &\ \text{(i)}\\
		\alpha_i\eta_i^2\sqrt{m'm} & \ \text{(ii)}\\
		-\frac{1}{2}\alpha_i\eta_i^2\sqrt{m'(m'-1)} & \ \text{(iii)},
	\end{cases}
	\label{eq:Rnmmn}
\end{align}
where $m'=\max(m_i',n_i')$, $m=\max(m_i,n_i)$. The numerical coefficients
$\alpha_j=\frac{1}{5}(2-\abs{\hat{d}_{eg,j}}^2)$ with $\frac{1}{5}\le \alpha_j\le
\frac{2}{5}$ and $\sum_j \alpha_j=1$ depend on the projection of the transitions
dipole matrix element onto the $j$-axis.

The total initial decay rate $\Gamma_{\text{eff}}$ defined in Eq.~\eqref{eq:init_decay} can
be written as
\begin{align*}
	\Gamma_{\text{eff}}&=\Gamma\left(
	1-R_{\boldsymbol{0}\boldsymbol{0}\boldsymbol{0}\boldsymbol{0}} \right),
\end{align*}
which in the general case can be evaluated numerically. For the case of tight trapping in
the Lamb-Dicke limit, by using Eq.~\eqref{eq:Rnmmn}, it is given by $\Gamma_{\text{eff}}=\sum_{j=1,2,3}\alpha_j
\eta_j^2\Gamma+\mathcal{O}(\eta_j^4)$.
\section{Weisskopf-Wigner approach}\label{sec:weisskopf-wigner}
For times $t\gg \Gamma_{\text{eff}}^{-1}$, the solution of the Weisskopf-Wigner ansatz
Eq.~\eqref{eq:weisskopf-wigner} in terms of the coefficients $a_{01,\boldsymbol{k}}(t)$,
$a_{0n,\boldsymbol{k}}(t)$, $a_{1n,\boldsymbol{k}}(t)$ with $n>1$  is \cite{Louisell1973} 
\begin{widetext}
	\begin{align*}
		a_{01,\boldsymbol{k}}(t)&=\mathcal{E}_k^\ast \boldsymbol{d}_{eg}\cdot
		\boldsymbol{e}_{\boldsymbol{k}}\mathrm{e}^{-\mathrm{i}(\omega_k+\nu)t}\left( 
		\frac{\mu_0R_{01}^\ast(\hat{\boldsymbol{k}}\cdot
		\hat{\boldsymbol{\chi}})}{\hbar\left( (\omega_k+\nu-\omega_0)+\mathrm{i}
		\Gamma^{(0)}_{\text{eff}}/2 \right)}-\frac{\mu_1R_{00}^\ast(\hat{\boldsymbol{k}}\cdot
		\hat{\boldsymbol{\chi}})}{\hbar \left( (\omega_k-\omega_0)+\mathrm{i}
		\Gamma^{(1)}_{\text{eff}}/2 \right)}
		\right)\displaybreak[0]\\
	a_{0n,\boldsymbol{k}}(t)&=\mathcal{E}_k^\ast \boldsymbol{d}_{eg}\cdot
	\boldsymbol{e}_{\boldsymbol{k}}\mathrm{e}^{-\mathrm{i}(\omega_k+n\nu)t}
	\frac{\mu_0R_{0n}^\ast(\hat{\boldsymbol{k}}\cdot
	\hat{\boldsymbol{\chi}})}{\hbar \left( (\omega_k+n\nu-\omega_0)+\mathrm{i}
	\Gamma^{(0)}_{\text{eff}}/2 \right)}\displaybreak[0]\\
	a_{1n,\boldsymbol{k}}(t)&=\mathcal{E}_k^\ast \boldsymbol{d}_{eg}\cdot
	\boldsymbol{e}_{\boldsymbol{k}}\mathrm{e}^{-\mathrm{i}(\omega_k+(n+1)\nu)t}\frac{\mu_1R_{0n}^\ast
	(\hat{\boldsymbol{k}}\cdot \hat{\boldsymbol{\chi}})}{\hbar\left(
	(\omega_k+(n+1)\nu-\omega_0)+\mathrm{i}\Gamma^{(1)}_{\text{eff}}/2 \right)}
		,
	\end{align*}
\end{widetext}
where $\Gamma^{(0)}_{\text{eff}}=\alpha\eta^2\Gamma+\mathcal{O}(\eta^4)$
($\Gamma^{(1)}_{\text{eff}}=(1-\alpha\eta^2)\Gamma+\mathcal{O}(\eta^4)$) 
is the effective decay rate for the initial state $\ket{g0;e0}$ ($\ket{g1;e0}$). All other
coefficients $a_{mn,\boldsymbol{k}}(t)$ with $n>m$ are zero within this model. 
The first order correlation
function $I(\boldsymbol{r},t)$ can be written as \cite{Scully1997}
\begin{widetext}
\begin{eqnarray*}
	I(\boldsymbol{r},t)&=&\abs{\boldsymbol{\Psi}_1(\boldsymbol{r},t)}^2+\sum_{n>1}\left[
	\abs{\boldsymbol{\Psi}_n^{(0)}(\boldsymbol{r},t)}^2+\abs{\boldsymbol{\Psi}_n^{(1)}(\boldsymbol{r},t)}^2
	\right]\\
	\abs{\boldsymbol{\Psi}_1(\boldsymbol{r},t)}^2&=&\frac{d_{eg}^2\omega_0^4 \sin^2
	\theta}{(4\pi\varepsilon_0)^2c^2r^2}\Theta(t-r/c)\notag\\
	&&\times \Bigl\{ \abs{\mu_0}^2 \abs{R_{01}(u)}^2 \exp\left(
	-\Gamma^{(0)}_{\text{eff}}(t-r/c) \right)+\abs{\mu_1}^2\abs{R_{00}(u)}^2\exp\left(
	-\Gamma^{(1)}_{\text{eff}}(t-r/c)
	\right)\Bigr.\notag\\
	&&\Bigl.-2\real\left( \mu_0 R_{01}^\ast(u)\mu_1^\ast R_{00}(u)\exp\left(
	\mathrm{i}\nu(t-r/c) \right)\exp\left(
	-(\Gamma^{(0)}_{\text{eff}}+\Gamma^{(1)}_{\text{eff}})(t-r/c)/2 \right) \right)
	\Bigr\}\\
	\abs{\boldsymbol{\Psi}_n^{(0,1)}(\boldsymbol{r},t)}^2&=&\frac{d_{eg}^2\omega_0^2\sin
	^2\theta \abs{\mu_{0,1}}^2\abs{R_{0n}(u)}^2}{(4\pi\varepsilon_0)^2c^2 r^2}\Theta(t-r/c)
	\exp\left( -\Gamma^{(0,1)}_{\text{eff}}(t-r/c) \right).
\end{eqnarray*}
\end{widetext}
where $\theta$ is the angle between
$\boldsymbol{r}$ and $\boldsymbol{d}_{eg}$, $u=\hat{\boldsymbol{r}}\cdot \hat{\boldsymbol{\chi}}$ 
and $\Theta$ is the Heaviside step function.
\bibliography{qo}
\end{document}